\documentclass[11pt,draftcls,journal,letterpaper,onecolumn]{IEEEtran}

\usepackage{graphicx}
\usepackage[cmex10]{amsmath}
\usepackage{amssymb}
\usepackage{bbm}
\usepackage{epstopdf}
\newcommand{\cA}{\mathcal{A}}
\newcommand{\cC}{\mathcal{C}}
\newcommand{\cL}{\mathcal{L}}
\newcommand{\cJ}{\mathcal{J}}

\newcommand{\cD}{\mathcal{D}}
\newcommand{\cR}{\mathcal{R}}
\newcommand{\cG}{\mathcal{G}}
\newcommand{\cN}{\mathcal{N}}

\newcommand{\ind}[1]{\mathbbm{1}_{\{#1\}}}     %indicator
\newcommand{\indz}[1]{\mathbbm{1}_{#1}}     %indicator

\newcommand{\bQ}{\mathbf{Q}}
\newcommand{\bI}{\mathbf{I}}
\newcommand{\bA}{\mathbf{A}}
\newcommand{\bB}{\mathbf{B}}
\newcommand{\bz}{\mathbf{0}}

\newcommand{\biQ}{\bar{\bQ}}

\newcommand{\iD}{{\mit\Delta}}
\newcommand{\iOm}{{\mit\Omega}}

\newcommand{\Exp}{{\sf E}}                  %Expectation
\newcommand{\Pro}{{\sf P}}                  %Probab
\newcommand{\Hyp}{{\sf H}}
\newcommand{\Dec}{{\sf D}}

\newcommand{\Real}{\text{Re}}

\newcommand{\Gauss}{\mathcal{N}_{\mathbb{C}}}
\newcommand{\hyptestoz}{\underset{\Hyp_0}{\overset{\Hyp_1}{\gtreqqless}}}
\newcommand{\hyptestozn}{\underset{\Hyp_0}{\overset{\Hyp_1}{\lesseqqgtr}}}
\newcommand{\hyptestru}{\underset{\Hyp_{1u}}{\overset{\Hyp_{1r}}{\lesseqqgtr}}}
\newcommand{\ignore}[1]{}

\newtheorem{theorem}{Theorem}
\newtheorem{lemma}{Lemma}

\begin{document}

\title{Joint Detection and Estimation: Optimum Tests and Applications}

%Article---------------------------------------------------------------
\author{George V.~Moustakides,~\IEEEmembership{Senior Member,~IEEE}, Guido~H.~Jajamovich,~\IEEEmembership{Student Member,~IEEE}, Ali~Tajer,~\IEEEmembership{Member,~IEEE} and Xiaodong Wang~\IEEEmembership{Fellow,~IEEE}
\thanks{Manuscript received\hskip2cm; revised\hskip2cm.}
\thanks{G.~V.~Moustakides is with the Electrical and
Computer Engineering Department, University of Patras, 26500 Rion, Greece, (e-mail: moustaki@upatras.gr).}
\thanks{G.~H.~Jajamovich and X.~Wang are with the Electrical Engineering Department, Columbia University, New York, NY 10027, USA, (e-mail: guido@ee.columbia.edu; wangx@ee.columbia.edu).}
\thanks{A.~Tajer is with the Electrical Engineering Department, Princeton University, Princeton, NJ 08544, USA, (e-mail: tajer@princeton.edu).}
}
%Digital Object Identifier\hskip1cm.
%\pubid{0000--0000/00\$00.00~\copyright~2003 IEEE}

\maketitle

\begin{abstract}
We consider a well defined joint detection and parameter estimation problem. By combining the Baysian formulation of the estimation subproblem with suitable constraints on the detection subproblem we develop optimum one- and two-step test for the joint detection/estimation case. The proposed combined strategies have the very desirable characteristic to allow for the trade-off between detection power and estimation efficiency. Our theoretical developments are then applied to the problems of retrospective changepoint detection and MIMO radar. In the former case we are interested in detecting a change in the statistics of a set of available data and provide an estimate for the time of change, while in the latter in detecting a target and estimating its location. Intense simulations demonstrate that by using the jointly optimum schemes, we can experience significant improvement in estimation quality with small sacrifice in detection power.
%\vskip0.5cm\noindent
%{\bf EDICS:} SSP-PARE, SSP-DETC, MSP-APPL
\end{abstract}


\markboth{IEEE Transactions on Information Theory (submitted)}{Moustakides et al.: Optimum joint detection and estimation: application to MIMO radar}

\section{Introduction}\label{sec:introduction}
\IEEEPARstart{T}here are important applications in practice where one is confronted with the problem of distinguishing between different hypotheses and, depending on the decision, to proceed and estimate a set of relevant parameters. Characteristic examples are: Detection and estimation of objects from images \cite{Vo}; Retrospective changepoint detection, where one desires to detect a change in statistics but also estimate the time of the change \cite{Boutoille,Vexler}; Defect detection from radiographies, where in addition to detecting presence of defects one would also like to find their position and shape \cite{Fillatre}; finally MIMO radar where we are interested in detecting the presence of a target and also estimate several target characteristics as position, speed, etc. All these applications clearly demand for detection and estimation strategies that address the two subproblems in a \textit{jointly} optimum manner.

In the literature, there are basically two (mainly ad-hoc) approaches that deal with combined problems. The first
consists in treating the two subproblems separately and applying in each case the corresponding optimum technique. For instance one can use the Neyman-Pearson optimum test for detection and the optimum Bayesian estimator for parameter estimation to solve the combined problem. As we will see in our analysis, and it is usually the case in combined problems, treating each part separately with the optimum scheme, does not necessarily result in optimum overall performance. The second methodology consists in using the Generalized Likelihood Ratio Test (GLRT) which detects and estimates at the same time with the parameter estimation part relying on the maximum likelihood estimator. Both approaches lack versatility and are not capable of emphasizing each subproblem according to the needs of the corresponding application.

Surprisingly, one can find \textit{very} limited literature that deals with optimum solutions of the joint detection and estimation problem. A purely Bayesian technique is reported in \cite{Middleton}, whereas a combination of Bayesian and Neyman-Pearson-like methodology is developed in \cite{Moustakides}. Specifically in \cite{Moustakides} the error probabilities under the two hypotheses, used in the classical Neyman-Pearson approach, are replaced by estimation costs. Mimicking the Neyman-Pearson formulation and constraining the estimation cost under the nominal hypothesis while optimizing the corresponding cost under the alternative, gives rise to a number of interesting combined tests that can be used in place of GLRT. 

Here we will build upon the methodology of \cite{Moustakides} but we are going to formulate the combined problem in a more natural way. In particular we will define a performance measure for the estimation part which we are going to optimize assuring, in parallel, the satisfactory performance of the detection part by imposing suitable constraints on the decision error probabilities. This idea will lead to two novel combined tests that have no equivalent in \cite{Middleton},\cite{Moustakides}.

We would like to point out that the theory in \cite{Middleton},\cite{Moustakides} as well as the one we are going to develop in our work, makes sense only when \textit{both} subproblems constitute desired goals in our setup, that is, when we are interested in detecting \textit{and} estimating. These results cannot provide optimum schemes for the case where one is interested \textit{only in detection} and is forced to use parameter estimation due to presence of nuisance parameters.

Our article is organized as follows: in Section\,II we define the joint detection and estimation problem and propose two different optimal solutions. As a quick example, our results are then applied to the problem of retrospective change detection. In Section\,III we make a thorough presentation of the MIMO radar problem under a joint detection and estimation formulation and use the results of the previous section in order to solve this problem optimally. Specifically we develop closed form expressions for all quantities that are needed to apply our theory and perform simulations to evaluate the performance of the optimum schemes, addressing also computational issues. Finally, in Section\,IV we have our concluding remarks.

\section{Optimum Joint Detection and Parameter Estimation}
Let us define the problem of interest. Motivated by most applications mentioned in the Introduction, we limit ourselves to the binary hypothesis case with parameters present only under the alternative hypothesis. Suppose we are given an observation signal $X$ for which we have the following two hypotheses

\indent$\Hyp_0:~X\sim f_0(X)$\\
\indent$\Hyp_1:~X\sim f_1(X|\theta),~\theta\sim\pi(\theta)$,

\noindent where $f_0(X),f_1(X|\theta),\pi(\theta)$ are known pdfs. Specifically, we assume that under $\Hyp_0$ we know the pdf of $X$ completely, whereas under $\Hyp_1$ the pdf of $X$ contains a collection of random parameters $\theta$ for which we have available some prior pdf $\pi(\theta)$. The goal is to develop a mechanism that distinguishes between $\Hyp_0,\Hyp_1$ and, furthermore, every time it decides in favor of $\Hyp_1$ it provides an estimate $\hat{\theta}$ for $\theta$. Our combined detection/estimation scheme is therefore comprised of a randomized test $\{\delta_0(X),\delta_1(X)\}$ with $\delta_i(X)$ denoting the randomization probability for deciding in favor of $\Hyp_i$; and a function $\hat{\theta}(X)$ that provides the necessary parameter estimates. Clearly $\delta_i(X)\ge0$ and $\delta_0(X)+\delta_1(X)=1$.

Let us recall, very briefly, the optimum detection and estimation theory when the two subproblems are considered separately.

\textit{Neyman-Pearson hypothesis testing}: Fix a level $\alpha\in(0,1)$; if $\Dec$ denotes our decision then we are interested in selecting a test (namely the randomization probabilities $\delta_i(X)$) so that the detection probability $\Pro_1(\Dec=\Hyp_1)$ is maximized subject to the false alarm constraint $\Pro_0(\Dec=\Hyp_1)\le\alpha$. Equivalently, the previous maximization can be replaced by the minimization of the probability of miss $\Pro_1(\Dec=\Hyp_0)$. The optimum detection scheme is the well celebrated likelihood ratio test, which takes the following form for our specific setup
\begin{equation}
\cL(X)=\frac{f_1(X)}{f_0(X)}=\frac{\int f_1(X|\theta)\pi(\theta)\,d\theta}{f_0(X)}\hyptestoz \gamma_{\text{NP}}.
\label{NP_test}
\end{equation}
In other words we decide $\Hyp_1$ whenever the likelihood ratio $\cL(X)$ exceeds the threshold $\gamma_{\text{NP}}$; $\Hyp_0$ whenever it falls below and randomize with a probability ${p}$ when the likelihood ratio is equal to the threshold. The threshold $\gamma_{\text{NP}}$ and the probability ${p}$ are selected to satisfy the false alarm constraint with equality.
The randomization probabilities $\delta_0^{\text{NP}}(X),\delta_1^{\text{NP}}(X)$ corresponding to the Neyman-Pearson test are given by
\begin{align}
\begin{split}
\delta_0^{\text{NP}}(X)&=\ind{\frac{f_1(X)}{f_0(X)}<\gamma_{\text{NP}}}+(1-{p})\ind{\frac{f_1(X)}{f_0(X)}=\gamma_{\text{NP}}}\\
\delta_1^{\text{NP}}(X)&=\ind{\frac{f_1(X)}{f_0(X)}>\gamma_{\text{NP}}}+{p}\ind{\frac{f_1(X)}{f_0(X)}=\gamma_{\text{NP}}},
\end{split}
\label{NP_deltas}
\end{align}
where $\indz{\cA}$ denotes the index function of the set $\cA$.

\textit{Bayesian parameter estimation}: Suppose that we know \textit{with certainty} that the observations $X$ come from hypothesis $\Hyp_1$, then we are interested in providing an estimate $\hat{\theta}(X)$ for the parameters $\theta$. We measure the quality of our estimate with the help of a cost function $C(\hat{\theta},\theta)\ge0$. We would like to select the optimum estimator in order to minimize the average cost $\Exp_1[C(\hat{\theta}(X),\theta)]$, where expectation is with respect to $X$ and $\theta$.

From \cite[Page 142]{Poor:book} we have that the optimum Bayesian estimator is the following minimizer (provided it exists)
\begin{equation}
\hat{\theta}_o(X)=\text{arg}\inf_U\cC(U|X),
\label{B_est1}
\end{equation}
where $\cC(U|X)$ is the posterior cost function
\begin{multline}
 \hfill
\cC(U|X)=\Exp_1[C(U,\theta)|X]=\frac{\int C(U,\theta)f_1(X|\theta)\pi(\theta)\,d\theta}{\int f_1(X|\theta)\pi(\theta)\,d\theta}  %\\%%
=\frac{\int C(U,\theta)f_1(X|\theta)\pi(\theta)\,d\theta}{f_1(X)},
 \hfill
\label{B_est2}
\end{multline}
and expectation, as we can see from the last equality, is with respect to $\theta$ for given $X$. Finally we denote the optimum posterior cost as $\cC_o(X)$, that is,
\begin{equation}
\cC_o(X)=\inf_U\cC(U|X)=\cC(\hat{\theta}_o(X)|X).
\label{o_cost}
\end{equation}
This quantity will play a very important role in the development of our theory as it constitutes a genuine quality index for the estimate $\hat{\theta}_o(X)$.

Let us now consider the combined problem. We recall that the hypothesis testing part distinguishes between $\Hyp_0$ and $\Hyp_1$. As we have seen, the Neyman-Pearson approach provides the best possible detection structure for controlling and optimizing the corresponding decision error probabilities. However with a decision mechanism that focuses on the decision errors, we cannot necessarily guarantee efficiency for the estimation part. Consequently, we understand, that the detection part cannot be treated independently from the estimation part. Following this rationale, we propose two possible approaches involving single and two-step schemes that differ in the number of decision mechanisms they incorporate and the way they combine the notion of \textit{reliable estimate} with the detection subproblem.

\subsection{Single-Step Tests}\label{sec:test1}
Let us begin our analysis by introducing a proper performance measure for the estimation subproblem. Following the Bayesian approach we assume the existence of the cost function $C(\hat{\theta},\theta)\ge0$. Computing the average cost that will play the role of our performance measure, is not as straightforward as in the pure estimation problem and requires some consideration. Note that an estimate $\hat{\theta}(X)$ is provided only when we decide in favor of $\Hyp_1$. On the other hand averaging of $C(\hat{\theta},\theta)$ makes sense only under the alternative hypothesis $\Hyp_1$ since under the nominal $\Hyp_0$ there is no true parameter $\theta$. Consequently we propose the following performance criterion
\begin{multline}
 \hfill
\cJ(\delta_0,\delta_1,\hat{\theta})=\Exp_1[C(\hat{\theta}(X),\theta)|\Dec=\Hyp_1]  %\\%%
=\frac{\Exp_1[C(\hat{\theta}(X),\theta)\ind{\Dec=\Hyp_1}]}{\Pro_1(\Dec=\Hyp_1)},
 \hfill
\label{cost}
\end{multline}
where expectation is with respect to $X$ and $\theta$. We realize that with our criterion, the estimation performance depends not only on the estimator but also on the detection mechanism. As we can see, we compute the average cost over the event $\{\Dec=\Hyp_1\}$, which is the only case an estimate is available. 

One would immediately argue that the measure in \eqref{cost} does not consider in any sense the decision errors, that is, the quality of the detector. However, these errors can be efficiently controlled through suitable constraints. Specifically we can impose the familiar false alarm constraint $\Pro_0(\Dec=\Hyp_1)\le\alpha$ but \textit{also} a constraint on the probability of miss $\Pro_1(\Dec=\Hyp_0)\le\beta$ where $\alpha,\beta\in(0,1)$. With these two constraints we have complete control over the decision mechanism and therefore, now, it makes sense to attempt to minimize the conditional average estimation cost $\cJ(\delta_0,\delta_1,\hat{\theta})$ over the decision rule $\{\delta_0(X),\delta_1(X)\}$ and the estimator $\hat{\theta}(X)$. Note that the two constraints guarantee satisfactory performance for the detection part and, by minimizing the criterion, we can enjoy optimum performance in the estimation part.

Let us carry out the desired optimization gradually. We first fix the decision rule $\{\delta_0(X),\delta_1(X)\}$ and optimize $\cJ(\delta_0,\delta_1,\hat{\theta})$ with respect to the estimator $\hat{\theta}(X)$. We have the following lemma that provides the solution to this problem.

\begin{lemma}\label{lem:1}
Let $\varphi(X)\ge0$ be a scalar function, then the following functional of $\hat{\theta}(X)$
\begin{equation}
\cD(\hat{\theta})=\frac{\iint \varphi(X)C(\hat{\theta}(X),\theta)f_1(X|\theta)\pi(\theta)d\theta dX}{\iint\varphi(X)f_1(X|\theta)\pi(\theta)d\theta dX}
\end{equation}
is minimized when $\hat{\theta}(X)$ is the optimum Bayesian estimator $\hat{\theta}_o(X)$ defined in \eqref{B_est1} and \eqref{B_est2}.
\end{lemma}
\begin{IEEEproof}
The proof is simple. We can write
\begin{multline}
\cD(\hat{\theta})=\frac{\iint \varphi(X)C(\hat{\theta}(X),\theta)f_1(X|\theta)\pi(\theta)d\theta dX}{\iint\varphi(X)f_1(X|\theta)\pi(\theta)d\theta dX}\\
=\frac{\int\varphi(X)\left(\int C(\hat{\theta}(X),\theta)f_1(X|\theta)\pi(\theta)d\theta\right) dX}{\int\varphi(X)\left(\int f_1(X|\theta)\pi(\theta)d\theta\right) dX}\\
=\frac{\int\varphi(X)\cC(\hat{\theta}(X)|X)f_1(X) dX}{\int\varphi(X)f_1(X)dX}\\
\ge\frac{\int\varphi(X)\inf_U\cC(U|X)f_1(X) dX}{\int\varphi(X)f_1(X)dX}\\
=\frac{\int\varphi(X)\cC(\hat{\theta}_o(X)|X)f_1(X) dX}{\int\varphi(X)f_1(X)dX}%\\%%
=\frac{\int\varphi(X)\cC_o(X)f_1(X) dX}{\int\varphi(X)f_1(X)dX},
\end{multline}
where for the last two equalities we used \eqref{o_cost}.
\end{IEEEproof}

Lemma\,\ref{lem:1} is a very interesting result because it demonstrates an extended optimality property for the classical Bayesian estimator. In particular by selecting $\varphi(X)=\delta_1(X)$ we conclude that $\hat{\theta}_o(X)$ continues to be optimum even if estimation is dictated by a decision mechanism and not performed over all data $X$, as is the usual practice with Bayesian estimation. Consequently, we can now fix our estimator to the Bayesian estimator $\hat{\theta}_o(X)$ with corresponding optimized performance measure equal to
\begin{equation}
\bar{\cJ}(\delta_0,\delta_1)=\cJ(\delta_0,\delta_1,\hat{\theta}_o)=\frac{\int\delta_1(X)\cC_o(X)f_1(X) dX}{\int\delta_1(X)f_1(X)dX}.
\label{J_cost}
\end{equation}
It is clear that our intention is to further minimize $\bar{\cJ}(\delta_0,\delta_1)$ over the class of detectors that satisfy the two error constraints. Before addressing this problem however, we need to make some remarks.

\textit{Remark 1:} One can argue that by constraining the false alarm probability to $\alpha$ and by using the Neyman-Pearson optimum test for detection and then the Bayesian estimator for estimation (in other words, treating the two subproblems separately) has definite optimality properties, since this combination optimizes both the detection and the estimation part. This is indeed true, however with such a scheme the main emphasis is on the detection part. For estimation, after optimizing the corresponding performance (by using $\hat{\theta}_o(X)$), we have no further control. In fact if the resulting estimation performance is not satisfactory, there is no room for further improvement. This weakness is clearly circumvented by the proposed formulation which offers, as we discuss next, the additional flexibility to trade detection power for estimation efficiency, according to the needs of the designer.

\textit{Remark 2:} We recall that in our setup we have the two constraints $\Pro_0(\Dec=\Hyp_1)\le\alpha$ and $\Pro_1(\Dec=\Hyp_0)\le\beta$. By fixing the false alarm probability to $\alpha$, the probability of miss is minimized by the Neyman-Pearson test; call this minimum value $\beta(\alpha)$. Since no test, with false alarm probability not exceeding $\alpha$, can have a probability of miss that goes below $\beta(\alpha)$, this suggests that in our constraint on the probability of miss, $\beta$ must be selected to satisfy $\beta\ge\beta(\alpha)$. We are thus reducing, in a controlled manner, the detection power as compared to the Neyman-Pearson test (since we allow more misses) aiming in improving the effectiveness of our estimation. We have the following theorem that provides the optimum scheme.

\begin{theorem}\label{th:1}
Consider the two constraints $\Pro_0(\Dec=\Hyp_1)\le\alpha$ and $\Pro_1(\Dec=\Hyp_0)\le\beta$, where $0<\alpha<1$ and $\beta(\alpha)\le\beta<1$ with $\beta(\alpha)$ denoting the probability of miss of the Neyman-Pearson test. Let $\lambda_o>0$ be the solution of the equation\footnote{For simplicity we assume that $\cC_o(X)$ and $f_1(X)/f_0(X)$, when considered as random variables, have no atoms under both hypotheses (the corresponding pdfs have no delta functions). This avoids the need for randomization every time a test statistic hits a threshold.}
\begin{equation}
\Pro_1\left(\lambda_o\ge\cC_o(X)\right)=1-\beta,
\label{eq:th1.1}
\end{equation}
where $\cC_o(X)$ is defined in \eqref{o_cost}.
Then the optimum combined scheme is comprised of the Bayesian estimator $\hat{\theta}_o(X)$ defined in \eqref{B_est1}, \eqref{B_est2}, for the estimation part while the decision rule that optimizes the average conditional cost $\bar{\cJ}(\delta_0,\delta_1)$ in \eqref{J_cost} under the two error constraints is given by
\begin{align}
\cC_o(X)&\hyptestozn \lambda_o,~~\text{if}~\alpha\ge\Pro_0\left(\lambda_o\ge\cC_o(X)\right)\label{test1a}
\\
\frac{f_1(X)}{f_0(X)}[\lambda-\cC_o(X)]&\hyptestoz \gamma,~~\text{if}~\alpha<\Pro_0\left(\lambda_o\ge\cC_o(X)\right),
\label{test1b}
\end{align}
where in \eqref{test1b} $\lambda,\gamma$ are selected so that the two error probability constraints are satisfied with equality.
\end{theorem}
\begin{IEEEproof}
The proof is presented in the Appendix.
\end{IEEEproof}

From \eqref{test1a} and \eqref{test1b} we deduce that the optimum detector takes into account the estimation part through $\cC_o(X)$ which constitutes a quality index for the estimate $\hat{\theta}_o(X)$. If this index is sufficiently large then, in both cases, the test decides in favor of $\Hyp_0$. In particular, in \eqref{test1b}, this decision may occur even if the classical likelihood ratio exceeds the threshold $\gamma_{\text{NP}}$, suggesting decision in favor of $\Hyp_1$.

Sumarizing, our first optimum combined test consist in applying \eqref{test1a} or \eqref{test1b} to decide between the two hypotheses and every time we make a decision in favor of $\Hyp_1$ we use $\hat{\theta}_o(X)$ defined in \eqref{B_est1} to provide the optimum parameter estimate.

\subsection{Two-Step Tests}\label{sec:test2}
In the previous setup our decision was between $\Hyp_0$ and $\Hyp_1$ and we were sacrificing detection power to improve estimation. However, in most applications, giving up part of the detection capacity may be regarded as undesirable. For example in MIMO radar it is still helpful to detect a target even if we cannot reliably estimate its parameters. 

It is possible to \textit{preserve} the detection power and at the same time \textit{ameliorate} the estimation performance if we follow a slightly different approach that involves \textit{two-step mechanisms}. Specifically we propose the use of an initial detection strategy that distinguishes between $\Hyp_0$ and $\Hyp_1$; whenever we decide in favor of $\Hyp_1$ then, at a second step, we compute the estimate $\hat{\theta}(X)$ and employ a \textit{second test} that decides whether the estimate is \textit{reliable} or \textit{unreliable}, denoted as $\Hyp_{1r}$ and $\Hyp_{1u}$ respectively. Consequently we propose to make \textit{three} different decisions $\Hyp_0,\Hyp_{1r}$ and $\Hyp_{1u}$ with the union of the last two corresponding to hypothesis $\Hyp_1$. As we can see, we ``trust'' the estimate $\hat{\theta}(X)$ only when we decide in favor of $\Hyp_{1r}$, but we have detection even if we discard the estimate as unreliable, that is, we decide $\Hyp_{1u}$.

For the first test we use our familiar randomization probabilities $\{\delta_0(X),\delta_1(X)\}$ while for the second we employ a new pair $\{q_{1r}(X),q_{1u}(X)\}$. The latter functions are the randomization probabilities needed to decide between reliable/unreliable estimation \textit{given} that the first test decided in favor of $\Hyp_1$. Therefore we have $q_{1r}(X),q_{1u}(X)\ge0$ and $q_{1r}(X)+q_{1u}(X)=1$. For every combination of the four randomization probabilities we define, similarly to the previous subsection, the corresponding average conditional cost for the estimator $\hat{\theta}(X)$, namely
\begin{multline}
 \hfill
\cJ(\delta_0,\delta_1,q_{1r},q_{1u},\hat{\theta})=\Exp_1[C(\hat{\theta}(X),\theta)|\Dec=\Hyp_{1r}]%\\%%
=\frac{\int \delta_1(X)q_{1r}(X)\cC(\hat{\theta}(X)|X)f_1(X)dX}{\int \delta_1(X)q_{1r}(X)f_1(X)dX}.
 \hfill
\end{multline}
As we can see, we now condition on the event $\{\Dec=\Hyp_{1r}\}$ since this is the only case when the estimate $\hat{\theta}(X)$ is accepted. We also note that, for given $X$, the probability to decide in favor of $\Hyp_{1r}$ is $\delta_1(X)q_{1r}(X)$ because we must decide in favor of $\Hyp_1$ in the first step (with probability $\delta_1(X)$) and for $\Hyp_{1r}$ in the second (with probability $q_{1r}(X)$).

In the first step we would like to adopt the best possible detector to select between $\Hyp_0$ and $\Hyp_1$. We follow the classical Neyman-Pearson approach and impose the false alarm probability constraint $\Pro_0(\Dec=\Hyp_1)\le\alpha$ while we minimize the probability of miss $\Pro_1(\Dec=\Hyp_0)$. This leads to the Neyman-Pearson test defined in \eqref{NP_test} with corresponding randomization probabilities $\delta_0^{\text{NP}}(X),\delta_1^{\text{NP}}(X)$ given in \eqref{NP_deltas}. 

Having identified the first, let us proceed to the second step of our detection/estimation mechanism that involves parameter estimation and a second test that labels the estimate as reliable/unreliable. Consider the average conditional cost $\cJ(\delta_0^{\text{NP}},\delta_1^{\text{NP}},q_{1r},q_{1u},\hat{\theta})$, assume $q_{1r}(X),q_{1u}(X)$ fixed, then from Lemma\,\ref{lem:1} and by selecting $\varphi(X)=\delta_1^{\text{NP}}(X)q_{1r}(X)$, we conclude that this criterion is minimized when $\hat{\theta}(X)=\hat{\theta}_o(X)$, that is, again with the optimum Bayes estimator defined in \eqref{B_est1} and \eqref{B_est2}. Call
\begin{multline}
 \hfill
\bar{\cJ}(q_{1r},q_{1u})=\cJ(\delta_0^{\text{NP}},\delta_1^{\text{NP}},q_{1r},q_{1u},\hat{\theta}_o)%\\%%
=\frac{\int\delta_1^{\text{NP}}(X)q_{1r}(X)\cC_o(X)f_1(X) dX}{\int\delta_1^{\text{NP}}(X)q_{1r}(X)f_1(X)dX},
 \hfill
\label{o_cost2}
\end{multline}
the corresponding performance. It is then clear that we would like to minimize even further this criterion by selecting properly our second decision mechanism which is expressed with the help of the randomization probabilities $\{q_{1r}(X),q_{1u}(X)\}$. Note however that, in addition to this minimization, we are also interested in generating as many ``reliable estimates'' as possible when applying the second test. These two goals are clearly conflicting, therefore we adopt a Neyman-Pearson-like approach in order to come up with an optimum scheme. In other words we constrain one quantity and optimize the other.

To find a suitable constraint, because $q_{1r}(X)\le1$, the probability $\Pro_1(\Dec=\Hyp_{1r})$ of deciding in favor of $\Hyp_{1r}$ (reliable estimate) satisfies
\begin{multline}
 \hfill
\Pro_1(\Dec=\Hyp_{1r})=\int\delta_1^{\text{NP}}(X)q_{1r}(X)f_1(X)dX  %\\%%
\le\int\delta_1^{\text{NP}}(X)f_1(X)dX=\Pro_1(\Dec=\Hyp_{1})=1-\beta(\alpha).
 \hfill
\end{multline}
In other words this probability is upper bounded by the detection probability $1-\beta(\alpha)$ of the Neyman-Pearson test where, we recall, $\beta(\alpha)$ denotes the corresponding probability of miss. This inequality reveals the obvious fact that, only a portion of our initial decisions in favor of $\Hyp_1$ provide reliable estimates in the second step. Actually it is this part we intend to control by imposing the following inequality
\begin{equation}
1-\beta\le\Pro_1(\Dec=\Hyp_{1r})=\int\delta_1^{\text{NP}}(X)q_{1r}(X)f_1(X)dX
\label{eq:2nd_test_con}
\end{equation}
with $1>\beta\ge\beta(\alpha)$. The constraint in \eqref{eq:2nd_test_con} expresses our desire that \textit{at least a fraction of $\frac{1-\beta}{1-\beta(\alpha)}\le\frac{\Pro_1(\Dec=\Hyp_{1r})}{\Pro_1(\Dec=\Hyp_{1})}\le1$ of the initial decisions in favor of $\Hyp_1$ must provide reliable estimates}. Subject to this constraint the goal is to obtain the best possible estimation performance, that is, minimize the performance measure $\bar{\cJ}(q_{1r},q_{1u})$. The solution to this optimization problem is given in the next lemma.

\begin{lemma}\label{lem:2}
Let $1>\beta\ge\beta(\alpha)$, then the test that minimizes the average conditional cost $\bar{\cJ}(q_{1r},q_{1u})$ defined in \eqref{o_cost2} subject to the constraint in \eqref{eq:2nd_test_con}, is given by
\begin{equation}
\cC_o(X)\hyptestru \lambda,
\label{eq:2_opt_test}
\end{equation}
where $\lambda$ is selected to satisfy \eqref{eq:2nd_test_con} with equality and $\cC_o(X)$ is defined in \eqref{o_cost}.
\end{lemma}
\begin{IEEEproof}
The proof follows a methodology which is very similar to the one used in the proof of Theorem\,\ref{th:1}. Since it presents no particular difficulties, it is omitted.
\end{IEEEproof}

As in the previous subsection, $\cC_o(X)$ constitutes a quality index for the estimate $\hat{\theta}_o(X)$. With Lemma\,\ref{lem:2} we end up with the very plausible decision rule of accepting $\hat{\theta}_o(X)$ as reliable whenever this index is below some threshold $\lambda$ while the estimate is discarded as unreliable whenever the same quantity exceeds the threshold. 

%Notice that, in the second combined scheme, the threshold $\gamma_{\text{NP}}$ is independent from the threshold $\lambda$ since it depends only on the false alarm probability $\alpha$. Once $\gamma_{\text{NP}}$ is specified we can then compute the second threshold $\lambda$ by satisfying the second constraint. This property is not enjoyed by the previous combined test where $\gamma$ and $\lambda$ are interrelated. Indeed in the single-step optimum scheme, if we change either of the two levels $\alpha,\beta$ this will affect \textit{both} parameters.

Summarizing our second detection/estimation scheme: We first use the Neyman-Pearson test \eqref{NP_test} to decide between $\Hyp_0,\Hyp_1$. Whenever we decide in favor of $\Hyp_1$ we compute the estimate $\hat{\theta}_o(X)$ from \eqref{B_est1} and its corresponding quality index $\cC_o(X)$ from \eqref{o_cost}; then we use the test in \eqref{eq:2_opt_test} to characterize the estimate as reliable/unreliable.

\subsection{MSE Cost and Uniform Prior}\label{ssec:2.C}
If we call $\cL(X|\theta)=\frac{f_1(X|\theta)}{f_0(X)}$ the conditional likelihood ratio, then all quantities entering in the two tests can be expressed with the help of $\cL(X|\theta)$ and the prior probability $\pi(\theta)$. We start with the likelihood ratio which is part of both tests and observe that we can write it as
\begin{equation}
\cL(X)=\frac{f_1(X)}{f_0(X)}=\int\cL(X|\theta)\pi(\theta)d\theta.
\label{eq:LR}
\end{equation}
From \eqref{B_est2} we can see that the posterior cost $\cC(U|X)$ can be computed as
\begin{equation}
\cC(U|X)=\frac{\int C(U,\theta)\cL(X|\theta)\pi(\theta)\,d\theta}{\int \cL(X|\theta)\pi(\theta)\,d\theta}
\label{eq:c_cost}
\end{equation}
suggesting that the Bayes estimator $\hat{\theta}_o(X)=\text{arg}\inf_U\cC(U|X)$ and the corresponding optimum posterior cost $\cC_o(X)=\inf_U\cC(U|X)$ can be expressed with the help of the conditional likelihood ratio as well.

Let us now examine the special case where for the cost function we adopt the squared error $C(U,\theta)=\|U-\theta\|^2$ which leads to the MSE criterion. From \cite[Page 143]{Poor:book}, we know that the optimum estimator $\hat{\theta}_o(X)$ is the conditional mean $\Exp_1[\theta|X]$. If we also assume the prior $\pi(\theta)$ to be uniform over some known set $\iOm$ with finite Lebesgue measure $\mu(\iOm)$ then
\begin{align}
\begin{split}
\cL(X)=\frac{f_1(X)}{f_0(X)}&=\mu^{-1}(\iOm)\int_{\iOm}\cL(X|\theta)d\theta\\
\hat{\theta}_o(X)&=\frac{\int_{\iOm}\theta \cL(X|\theta)d\theta}{\int_{\iOm}\cL(X|\theta)d\theta}\\
\cC_o(X)&=\frac{\int_{\iOm}\|\hat{\theta}_o(X)-\theta\|^2 \cL(X|\theta)d\theta}{\int_{\iOm}\cL(X|\theta)d\theta}\\
&=\frac{\int_{\iOm}\|\theta\|^2 \cL(X|\theta)d\theta}{\int_{\iOm}\cL(X|\theta)d\theta}-\|\hat{\theta}_o(X)\|^2.
\end{split}
\label{eq:final_test}
\end{align}
We can see that $\mu(\iOm)$ does not enter in the computation of the estimate $\hat{\theta}_o(X)$ and its quality index $\cC_o(X)$. Although $\mu(\iOm)$ does appear in the likelihood ratio $\cL(X)$, it is easy to verify that, in both tests, it can be transferred to the right hand side and absorbed by the corresponding threshold $\gamma$. We therefore conclude that no explicit knowledge of this quantity is necessary. Finally, we note that in the MSE criterion, $\cC_o(X)$ is the conditional variance of $\hat{\theta}_o(X)$ which clearly constitutes a very reasonable quality index for the corresponding estimate.

We have now completed the development of our theory that addresses the joint detection and estimation problem. To demonstrate the power and originality of our analysis, first we apply our results to the example of retrospective change detection and then in Section\,\ref{sec:model}, at a much greater extent, we use them to solve the MIMO radar problem.

\subsection{Example: Retrospective Change Detection}
Retrospective change detection is the problem where within a given set of data $X=[x_1,\ldots,x_N]$ there is a possible time instant $\tau$ where the data switch statistics from some nominal pdf $f(X)$ before $\tau$ to an alternative pdf $h(X)$ after $\tau$. We consider $\tau$ as the \textit{last} time instant under the nominal regime. 
Given $X$ we are interested in detecting the change but also estimating the time $\tau$ the change took place.

We should point out that retrospective change detection methodology is largely dominated by \textit{sequential} techniques \cite{Vexler}. However, this constitutes a serious misusage of these methods since, in the retrospective formulation, the data are all available at once, whereas in the sequential setup the data become available sequentially. This means that by adopting sequential tests for the solution of the retrospective problem results in an inefficient utilization of the existing information.

Let us now apply our previous theory. Note that for $0\le\tau<N$, the two pdfs can be decomposed as
\begin{align}
\begin{split}
f(X)&=f(x_1,\ldots,x_\tau)\times f(x_{\tau+1},\ldots,x_N|x_1,\ldots,x_\tau)\\
h(X)&=h(x_1,\ldots,x_\tau)\times h(x_{\tau+1},\ldots,x_N|x_1,\ldots,x_\tau).
\end{split}
\label{eq:decomp}
\end{align}
We first need to define the data pdf under the two hypotheses. Under $\Hyp_0$ we are under the nominal model therefore, clearly, $f_0(X)=f(X)$. Under $\Hyp_1$ and with a change occurring at $\tau$, we define the pdf $f_1(X|\tau)$ as follows
\begin{equation}
f_1(X|\tau)=f(x_1,\ldots,x_\tau)\times h(x_{\tau+1},\ldots,x_N|x_1,\ldots,x_\tau).
\end{equation}
In other words, from the decompositions in \eqref{eq:decomp}, we combine the first part of the nominal pdf with the second part of the alternative. With this changepoint model, the data before the change affect the data after the change through the conditional pdf. This is the most common model used in change detection theory \cite{Sasha}. Note that $\tau>N-1$ means that all the data are under the nominal regime (i.e.~there is no change) whereas $\tau=0$ that all the data are under the alternative regime. Therefore, under $\Hyp_1$ we have $\tau\in\{0,\ldots,N-1\}$ with some prior $\{\pi_0,\ldots,\pi_{N-1}\}$.

Let us compute the quantities that are necessary to apply our tests. Using \eqref{eq:decomp} we can write for the conditional likelihood ratio
\begin{equation}
\cL(X|\tau)=\frac{h(x_{\tau+1},\ldots,x_N|x_1,\ldots,x_\tau)}{f(x_{\tau+1},\ldots,x_N|x_1,\ldots,x_\tau)},
\end{equation}
suggesting that the likelihood ratio, from \eqref{eq:LR}, takes the form $\cL(X)=\sum_{\tau=0}^{N-1}\pi_{\tau}\cL(X|\tau)$.

Consider now the estimation problem. We propose the following cost function $C(U,\tau)=\ind{U\neq\tau}$, penalizing incorrect estimates by a unit cost. The average cost is clearly the probability to estimate incorrectly. Observing that $\ind{U\neq\tau}=1-\ind{U=\tau}$, from \eqref{eq:c_cost} we can write
\begin{equation}
\cC(U|X)=1-\frac{\cL(X|U)\pi_U}{\sum_{\tau=0}^{N-1}\pi_{\tau}\cL(X|\tau)}=1-\frac{\cL(X|U)\pi_U}{\cL(X)}.
\end{equation}
Consequently the optimum estimator that minimizes $\cC(U|X)$ over $U\in\{0,\ldots,N-1\}$ is
\begin{equation}
\hat{\tau}_o(X)=\text{arg}\max_{0\le U\le N-1}\cL(X|U)\pi_U,
\end{equation}
which is the MAP estimator \cite[Pages 145-150]{Poor:book}; while the corresponding optimum posterior cost becomes
\begin{equation}
\cC_o(X)=1-\frac{\max_{0\le U\le N-1}\cL(X|U)\pi_U}{\cL(X)}.
\end{equation}

The classical test that treats the two subproblems separately consists in comparing the likelihood ratio $\cL(X)$ to  the threshold $\gamma_{\text{NP}}$ in order to distinguish between the two hypotheses and use $\hat{\tau}_o(X)$ to estimate the time of change. GLRT on the other hand compares $\max_{0\le U\le N-1}\cL(X|U)$ to a threshold with the argument of this maximization providing the estimate for the time of change.

Applying our theory to this problem, for the single-step test we use $\hat{\tau}_o(X)$ for the estimate of the changetime and either 
\begin{equation}
\frac{\max_{0\le U\le N-1}\cL(X|U)\pi_U}{\cL(X)}\hyptestoz(1-\lambda_o),
\end{equation}
or
\begin{equation}
(\lambda-1)\cL(X)+\max_{0\le U\le N-1}\cL(X|U)\pi_U\hyptestoz \gamma,
\end{equation}
for the decision. For the two-step scheme we compare the likelihood ratio $\cL(X)$ to the threshold $\gamma_{\text{NP}}$ to decide between the two hypotheses; use $\hat{\tau}_o(X)$ for the changepoint estimate and finally apply
\begin{equation}
\frac{\max_{0\le U\le N-1}\cL(X|U)\pi_U}{\cL(X)}
{\underset{\Hyp_{1u}}{\overset{\Hyp_{1r}}{\gtreqqless}}}(1-\lambda)
\end{equation}
to label the estimate as reliable/unreliable.
Both combined schemes resulting from our theory, are completely original and make efficient use of all available information.

\section{Application to MIMO Radar}\label{sec:model}
A context where performing joint detection and estimation is of particular interest is in radar systems. Radars are often deployed not only to detect a target but also estimate unknown parameters associated with the target, e.g., position and velocity. Recent developments in radar systems equip radars with multiple transmit and receive arrays that considerably improve their detection power and estimation accuracy compared with the conventional phased-array radars.

In this section we examine the merits of the tests developed in the previous section for enhancing the detection and estimation quality by employing multiple-input multiple-output (MIMO) radar systems with widely-separated antennas~\cite{Haimovich:SPM08}. In particular we are interested in the detection of a target, and the estimation of its location every time a target is ruled present. This is somewhat different from the more conventional approaches in MIMO radar systems,  e.g., \cite{Fishler:SP06} and references therein, where the probe space is broken into small subspaces and the radar detects the presence of the target in each of the subspaces separately. In this approach as the location to be probed is given, one is only testing whether a target is present in a certain given subspace \cite{Fishler:SP06}. This necessitates implementing multiple detection tests in parallel, one for each subspace. In this section, we develop detectors and estimators based on the optimality theory discussed in the previous section which are used only once for the entire space.

\subsection{System Description}
We consider a MIMO radar system with $M$ transmit and $N$ receive antennas that are \textit{widely} separated (satisfy the conditions in \cite[Sec. II.A]{Fishler:SP06}). Such spacing among the antennas ensures that the receivers capture uncorrelated reflections from the target. Both transmit and receive antennas are located at positions $\theta^{t}_m\in\mathbb{R}^3$, for $m\in\{1,\dots,M\}$, and $\theta^{r}_n\in\mathbb{R}^3$, for $n\in\{1,\dots,N\}$, respectively, known at the receiver.

The $m$th transmit antenna emits the waveform with baseband equivalent model given by $\sqrt{E}s_m(t)$ where $E$ is the transmitted energy of a single transmit antenna (assuming to be the same for all transmitters); $\int_0^{T_s}|s_m(t)|^2dt=1$ and $T_s$ denotes the common duration of all signals $s_m(t)$.

We aim to detect the presence of an \emph{extended} target and when deemed to be present also estimate its position. The extended target consists of multiple scatterers exhibiting random, independent and isotropic scintillation, each modeled with a complex random variable of zero-mean and unknown distribution. This corresponds to the classical Swerling case I model \cite{Skolnik:book} extended for
multiple-antenna systems \cite{Haimovich:SPM08,Fishler:SP06}. The reflectivity factors are assumed to remain constant during a scan and are allowed to change independently from one scan to another.

We define $\theta$ as the location of the gravity center of the target and $d_{mn}(\theta)$ as the aggregate distance that a probing waveform $s_m(t)$ travels from the $m$th transmit antenna to the target and from the target to the $n$th receive antenna, i.e.,
\begin{equation}
  d_{mn}(\theta)= \sqrt{\|\theta-\theta_m^t\|^2_2+\|\theta-\theta_n^r\|^2_2}.
\label{eq:tau1}
\end{equation}
The time delay the waveform $s_m(t)$ is experiencing by traveling this distance $d_{mn}(\theta)$ is equal to
\begin{equation}
  \tau_{mn}(\theta)=\frac{d_{mn}(\theta)}{c},
\label{eq:tau2}
\end{equation}
where $c$ is the speed of light. When the target dimensions are considerably smaller than the distance of the target from the transmit and receive antennas, the distance of the antennas to each scatterer of the target can be well-approximated by their distances from the gravity center of the target. Therefore, the received signal
at the $n$th receive antenna is the superposition of all emitted
waveforms and is given by \cite{Tajer:STSP10}
\begin{equation}
    \label{eq:signalmodel} r_n(t) = \sqrt{E}\sum_{m=1}^{M}d_{mn}^{-\eta}(\theta)\;g_{mn} \;s_m(t-\tau_{mn}(\theta))+ w_{n}(t),
\end{equation}
where $d_{mn}^{-\eta}$ is the path-loss with $\eta$ denoting the path-loss exponent; $ w_n(t)$ the additive white Gaussian complex valued noise distributed as\footnote{$\Gauss(\mu,\sigma^2)$ denotes the distribution of a complex Gaussian random variable with mean $\mu=\mu_r+j\mu_i$ where the real and imaginary parts are uncorrelated (and therefore independent) Gaussian random variables with mean $\mu_r,\mu_i$ respectively and of variance equal to $\sigma^2/2$.} $\mathcal{N}_{\mathbb{C}}(0,1)$; and $g_{mn}$ accounts for the reflectivity effects of the scatterers corresponding to the $m$th transmit and the $n$th receive antennas. It can be readily verified that $\{g_{mn}\}$ are independent and identically distributed (i.i.d.) with distribution  $\mathcal{N}_{\mathbb{C}}(0,1)$ \cite{Fishler:SP06, Tajer:STSP10}. We note that we have assumed for the noises $w_n(t)$ and the coefficients $g_{mn}$ that they have variance equal to 1. In fact if we use any other values e.g.~$\sigma_w^2$ and $\sigma_g^2$ respectively then in the final test these quantities are combined with the transmitted signal power $E$ in the form of $E\sigma_g^2/\sigma_w^2$. Consequently, \textit{provided} that in the general case $\sigma_w^2$ and $\sigma_g^2$ are known then, without loss of generality, we may assume $\sigma_w^2=\sigma_g^2=1$ and let $E$ express the final combination.

For $n\in\{1,\dots,N\}$ define 
\begin{align}
\begin{split}
G_n^H&=[g_{1n},\hdots, g_{Mn}]\\
S_{n}'(t,\theta)&=\sqrt{E}\left[\frac{s_1(t-\tau_{1n}(\theta))}{d^\eta_{1n}(\theta)} ,\dots, \frac{s_M(t-\tau_{Mn}(\theta))}{d^\eta_{Mn}(\theta)}\right],
\end{split}
\end{align}
where we recall that $d_{mn}(\theta)$ and $\tau_{mn}(\theta)$ are \textit{known} functions of $\theta$ defined in \eqref{eq:tau1}, \eqref{eq:tau2} and $\bA',\bA^H$ denote the transpose and Hermitian (transpose and complex conjugate) respectively of the matrix $\bA$. Under these definitions we can write
\begin{equation}
 \label{eq:modelmatrixform}
    r_n(t)=  G_n^H  \cdot S_{n}(t,\theta) +  w_n(t).
\end{equation}
Let us now formulate the joint detection and estimation problem for the specific signal model we just introduced.

\subsection{Target Detection/Localization with MIMO Radar}
For $0\le t\le T$, we distinguish the following two hypotheses satisfied by the received signals $r_n(t),~n=1,\ldots,N$,

\indent$\Hyp_0:~dr_n(t)=dw_n(t)$\\
\indent$\Hyp_1:~dr_n(t)=G_n^H S_{n}(t,\theta)dt +  dw_n(t)$.

\noindent We have written the received signals in a stochastic differential equation form, since the $\{w_n(t)\}$ are Wiener (white Gaussian noise) processes. As we can see, when there is no target present the measured signals are pure Wiener processes, whereas with the appearance of a target we have the emergence of the nonzero drifts $G_n^H S_{n}(t,\theta)$. 

For simplicity, let us use $\bar{r}_n$ to denote the signal acquired by the $n$th receive antenna during the time-interval $[0,T]$, that is, $\bar{r}_n=\{r_n(t),~0\le t\le T\}$. The collection of these $N$ signals constitutes the complete set of observations, in other words, $\{\bar{r}_1,\ldots,\bar{r}_N\}$ plays the role of the observation signal $X$ of the previous section. 
Clearly, our goal is to use $\{\bar{r}_1,\ldots,\bar{r}_N\}$ in order to decide between the presence or absence of a target and, every time a target is detected, to provide a reliable estimate of its position.

To apply the theory developed in the previous section, according to Section\,\ref{ssec:2.C}, we need to find the conditional likelihood ratio $\cL(\bar{r}_1,\ldots,\bar{r}_N|\theta)$. The following theorem provides the required formula.

\begin{theorem}\label{th:2}
The likelihood ratio $\cL(\bar{r}_1,\ldots,\bar{r}_N|\theta)$ of the received signals is given by
\begin{equation}
\cL(\bar{r}_1,\ldots,\bar{r}_N|\theta)=\prod_{n=1}^N\frac{e^{R_n^H(\theta)(\bQ_n(\theta)+\bI_M)^{-1}R_n(\theta)}}{|\bQ_n(\theta)+\bI_M|},
\label{eq:th2.1}
\end{equation}
where
\begin{align}
\begin{split}
\bQ_n(\theta)&=\int_0^TS_n(t,\theta)S_n^H(t,\theta)dt,\\
R_n^H(\theta)&=\int_0^TS_n^H(t,\theta)dr_n(t),
\end{split}
\label{eq:th2.2}
\end{align}
$\bI_K$ denotes the identity matrix of size $K$ and $|\bA|$ the determinant of the matrix $\bA$.
\end{theorem}
\begin{IEEEproof}
The proof is presented in the Appendix.
\end{IEEEproof}

A final quantity that is of major interest for the next section is the appropriate definition of SNR. Note that, depending on the position of the target, the received signals $r_n(t)$ exhibit different SNR levels. This is due to the path-loss effect, which is particularly severe for distant targets. We therefore propose to measure the SNR by aggregating the signal and noise energies at the receivers but also \textit{averaging} these quantities over all possible target positions $\theta\in\iOm$. Specifically, by adopting the uniform model for $\theta$, we define
\begin{multline}
 \hfill
\text{SNR}=\frac{\int_{\iOm}\left(\sum_{n=1}^N\int_0^T\Exp[|G_n^HS_n(t,\theta)|^2]\,dt\right)d\theta}{\int_{\iOm}\left(\sum_{n=1}^N\Exp[|\int_0^Tdw_n(t)|^2]\right)d\theta}  %\\%%
\approx\frac{E}{N T}\frac{1}{\mu(\iOm)}\int_{\iOm}\left(\sum_{n=1}^N\sum_{m=1}^M\frac{1}{d_{mn}^{2\eta}(\theta)}\right)\,d\theta,
 \hfill
\label{eq:SNR}
\end{multline}
where from standard It\^o Calculus the expectation in the denominator is equal to $T$. For the approximate equality we overlooked the boundary effects in the numerator, that is, we assumed that $\int_0^T|s_n(t-\tau_{mn}(\theta))|^2dt=1$ for all $\tau_{mn}(\theta)$ which, of course, is not true when $\theta$ is close to the boundary of $\iOm$. If there is no path-loss, that is $\eta=0$, then the previous equation reduces to the simple formula $\text{SNR}\approx\frac{E\times M}{T}$. The transmitted energy $E$ will be tuned through these equations in order to deliver the appropriate SNR level at the receivers.

We have now developed all necessary formulas that enable us to use the results of Section\,II in the MIMO radar problem. In the next subsection we evaluate the joint detection/estimation scheme with Monte-Carlo simulations that cover various combinations of SNR values and number of transmit/receive antennas. We apply only the two-step test developed in Section\,\ref{sec:test2} since, as we briefly argued earlier, it is more well suited for the MIMO radar problem.

\subsection{Simulations}\label{sec:simu}
We consider the two-dimensional analog of the MIMO radar problem with two configurations consisting of $M=N=2$ and $M=N=3$ antennas, where the $m$th transmit and the $n$th receive antenna are located at $\theta^t_m=[m,0]'$ and $\theta^r_n=[0,n]'$ (expressed in Km), respectively.

The emitted waveforms are $s_m(t)=\frac{1}{\sqrt{T_s}}e^{j\frac{2\pi m}{T_s}t}$ for $t\in[0,T_s]$ where $T_s=10^{-4}$\,sec is the signal duration. Moreover, we select an integration time $T=5\times T_s=5\times 10^{-4}$\,sec. This integration limit can accommodate delays $\tau_{mn}(\theta)$ that do not exceed $T$ (for larger delays we simply measure noise during the interval $[0,T]$). The maximal delay defines a region $\iOm$ in space where every point $\theta\in\iOm$ has at least one aggregate distance $d_{mn}(\theta)$, defined in \eqref{eq:tau1}, from one transmit and one receive antenna that does not exceed the value $c\times T\approx 150$\,Km. Actually, the points in space that have an aggregate distance from a pair of transmit/receive antennas not exceeding 150\,Km lie in the interior of a well defined ellipse. Since we have $M\times N$ pairs of transmit/receive antennas, we conclude that $\iOm$ is the union of an equal number of such ellipses. By considering that all antennas are roughly positioned at the origin, all ellipses become circles and $\iOm$ can be approximated by a disc of approximate radius of 75\,Km. 

As is the usual practice in MIMO radar literature, we assume $\eta=0$, namely, no path-loss. This means that we are going to tune our energy parameter $E$ through the simplified equation $\text{SNR}\approx\frac{E\times M}{T}$. We consider SNR values -20, -10, 0 and 10\,dB.

Assuming that the target position $\theta$ is uniformly distributed within $\iOm$ and that for the cost function we employ the MSE criterion, we can use the formulas in \eqref{eq:final_test} for the joint detection/estimation scheme. From \eqref{eq:final_test} and \eqref{eq:th2.2} we observe the need for space and time integration. Both integrals will be evaluated numerically. For time integration we use canonical sampling and consider $L_t$ points $\{t_k\}$ within the time-interval $[0,T]$. For integration in space we form a canonical square grid of points for $\theta$. Denote with $L_s$ the number of points $\{\theta_l\}$ that lie in the interior of the region $\iOm$. The two integrals are then approximated by sums.
Specifically, the quantities in \eqref{eq:th2.2}, for $\theta=\theta_l$, are approximated by
\begin{align}
\bQ_n(\theta_l)\approx\frac{T}{L_t}\times\sum_{k=1}^{L_t}S_n(t_k,\theta_l)S_n^H(t_k,\theta_l)
\end{align}
and $R_n(\theta_l)$ under $\Hyp_0$ (needed to compute the threshold $\gamma_{\text{NP}}$) takes the form
\begin{equation}
R_n^H(\theta_l)\approx
\sum_{k=1}^{L_t}S_n^H(t_k,\theta_l)\iD w_n(t_k),
\end{equation}
while for the same quantity under $\Hyp_1$ we can write
\begin{equation}
R_n^H(\theta_l)\approx\sum_{k=1}^{L_t}S_n^H(t_k,\theta_l)\{G_n^H S_{n}(t_k,\theta_o)\times\frac{T}{L_t} +  \iD w_n(t_k)\}.
\end{equation}
Parameter $\theta_o$ denotes the ``true''  target position selected uniformly within $\iOm$ and $\theta_l$ is one of the $L_s$ grid-points in the interior of the same set. The coefficients $G_n$ are selected randomly from a Gaussian $\Gauss(0,\bI_M)$ while each $\iD w_n(t)$ is also Gaussian $\Gauss(0,\frac{T}{L_t})$. For each run, the quantities $G_n,\theta_o$ and $\iD w_n(t_k)$ are the same for all $\theta_l$. For our simulations we use $L_t=500$ time samples $\{t_k\}$ and a grid with cells 10\,Km$\times$10\,Km that generates 179 points $\{\theta_l\}$ in the interior of $\iOm$.

For the test of Section\,\ref{sec:test2}, according to \eqref{eq:final_test}, the likelihood ratio test is implemented as
\begin{equation}
\sum_{\theta_l}\prod_{n=1}^N\frac{e^{R_n^H(\theta_l)(\bQ_n(\theta_l)+\bI_M)^{-1}R_n(\theta_l)}}{|\bQ_n(\theta_l)+\bI_M|}\hyptestoz \gamma.
\end{equation}
Every time a decision is made in favor of $\Hyp_1$ we provide the following estimate of $\theta_o$
\begin{equation}
\hat{\theta}_o=\frac{\displaystyle\sum_{\theta_l}\theta_l\prod_{n=1}^N\frac{e^{R_n^H(\theta_l)(\bQ_n(\theta_l)+\bI_M)^{-1}R_n(\theta_l)}}{|\bQ_n(\theta_l)+\bI_M|}}{\displaystyle\sum_{\theta_l}\prod_{n=1}^N\frac{e^{R_n^H(\theta_l)(\bQ_n(\theta_l)+\bI_M)^{-1}R_n(\theta_l)}}{|\bQ_n(\theta_l)+\bI_M|}},
\end{equation}
with corresponding quality index
\begin{equation}
\cC_o=\frac{\displaystyle\sum_{\theta_l}\|\theta_l\|^2\prod_{n=1}^N\frac{e^{R_n^H(\theta_l)(\bQ_n(\theta_l)+\bI_M)^{-1}R_n(\theta_l)}}{|\bQ_n(\theta_l)+\bI_M|}}{\displaystyle\sum_{\theta_l}\prod_{n=1}^N\frac{e^{R_n^H(\theta_l)(\bQ_n(\theta_l)+\bI_M)^{-1}R_n(\theta_l)}}{|\bQ_n(\theta_l)+\bI_M|}}-\|\hat{\theta}_o\|^2.
\end{equation}
The estimate $\hat{\theta}_o$ is characterized as reliable/unreliable depending on whether $\cC_o$ is below/exceeds the threshold $\lambda$.

We also consider the GLRT where we maximize the likelihood ratio $\cL(\bar{r}_1,\ldots,\bar{r}_N|\theta)$ in \eqref{eq:th2.1} over $\theta$ and compare it to a threshold. The threshold is selected so that the corresponding false alarm probability is equal to $\alpha$. We recall that GLRT provides ML estimates for $\theta$ and, as we mentioned, cannot trade detection power for estimation.

Monte Carlo simulations were carried out in order to study the performance of the different tests. For each SNR value, 200,000 simulations were implemented to validate our theoretical developments. In our simulations we fixed the false alarm probability to $\alpha=10^{-3}$. The (conditional) MSE was computed as $\frac{1}{K}\sum \|\hat{\theta}_o-\theta_o\|^2$ where $K$ is the total number of cases where the combined test decided in favor of $\Hyp_{1r}$ (that is, $\Hyp_1$ in the first step and $\Hyp_{1r}$ in the second). 

\begin{figure}
\centering
\includegraphics{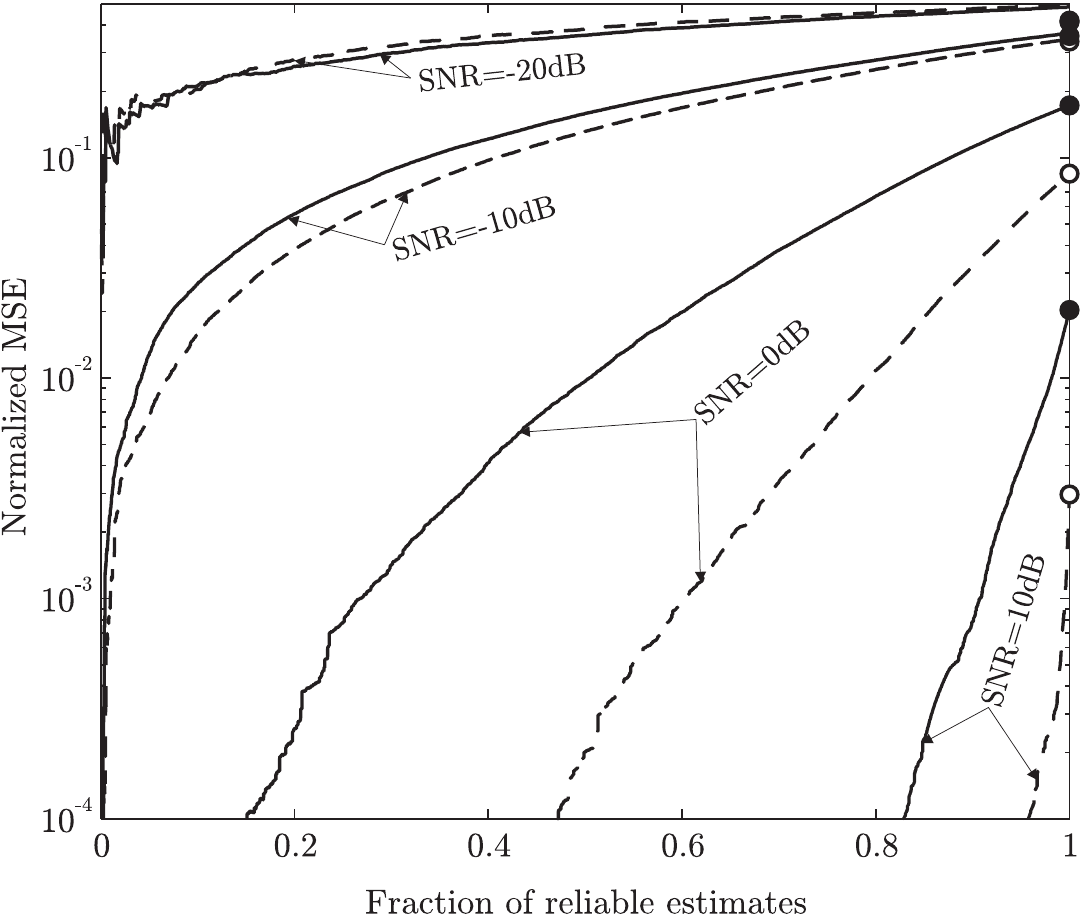}
\caption{Normalized MSE as a function of the fraction of reliable estimates for different values of SNR. Configuration $M=N=2$: optimum is solid and GLRT is $\bullet$; configuration $M=N=3$: optimum is dashed and GLRT is $\circ$.}
\label{fig:1}
\end{figure}
In Fig.\,\ref{fig:1} we depict the MSE normalized by the (approximate) radius of $\iOm$ squared ($75^2$) as a function of the fraction of reliable estimates, i.e.~$\frac{\Pro_1(\Dec=\Hyp_{1r})}{\Pro_1(\Dec=\Hyp_{1})}$. The fraction value is controlled through the threshold $\lambda$. Fraction value equal to 1 in our test corresponds to the performance of the classical approach where detection and estimation are treated separately. For the same value we also present the performance of the GLRT. We observe that for $\text{SNR}=-20$\,dB we need to sacrifice more than 50\% of our detections (more accurately in these cases we regard the estimates as unreliable) to reduce the MSE by a factor of 2. For larger SNR values we can have significant (even enormous) gains. For example for $\text{SNR}=0$\,dB by sacrificing 50\% of the detections, in the $2\times2$ case we gain an order of magnitude in estimation performance while the same gain in the $3\times3$ configuration is achieved with only 25\% reduction. We conclude from our simulations that apart the very low SNR case of $-20$\,dB, the $3\times3$ antenna configuration is preferable to the $2\times2$ since it can return significant performance gains. Finally, we observe that GLRT and the classical approach that treats the two subproblems separately have very comparable performance.

\section{Conclusion}
We have presented two possible formulations of the joint detection and estimation problem and developed the corresponding optimum solutions. Our approach consists in properly combining the Bayesian method for estimation with suitable constraints on the detection part. The resulting optimum schemes allow for the trade-off between detection power and estimation efficiency, thus emphasizing each subproblem according to needs of the original application. Our theory was then applied to the problems of retrospective change detection and MIMO radar. In particular in the second application, intense simulations demonstrated the possibility to experience significant gains in estimation quality with small sacrifices in detection power.

\section{Appendix}

{\em Proof of Theorem\,\ref{th:1}:} We are interested in minimizing $\bar{\cJ}(\delta_0,\delta_1)$ defined in \eqref{J_cost} subject to the two constraints $\int\delta_1(X)f_0(X)dX\le\alpha$ and $\int\delta_0(X)f_1(X)dX\le\beta$. We first note that if we have a pair $\{\delta_0(X),\delta_1(X)\}$ for which the second inequality is strict, then we can find another pair $\{\bar{\delta}_0(X),\bar{\delta}_1(X)\}$ which satisfies the second constraint with equality and has exactly the same estimation performance. Indeed if we select $
\bar{\delta}_1(X)=\frac{1-\beta}{\int\delta_1(X)f_1(X)dX}\delta_1(X)$, $\bar{\delta}_0(X)=1-\bar{\delta}_1(X)$,
then we observe that since we assumed $\int\delta_0(X)f_1(X)dX<\beta$ we have $\int\delta_1(X)f_1(X)dX=1-\int\delta_0(X)f_1(X)dX>1-\beta$, suggesting that $\bar{\delta}_1(X)$ is a legitimate probability (because $\delta_1(X)$ is multiplied by a factor smaller than 1 to produce $\bar{\delta}_1(X)$), consequently the complementary probability $\bar{\delta_0}(X)$ is legitimate as well. The fact that the alternative pair has exactly the same estimation performance, namely $\bar{\cJ}(\delta_0,\delta_1)=\bar{\cJ}(\bar{\delta}_0,\bar{\delta}_1)$, can be verified by direct substitution.

With the previous observation we can limit our search for the optimum within the class of tests that satisfy the constraint on the probability of miss with equality, that is, $\int\delta_0(X)f_1(X)dX=\beta$. Equivalently we consider only tests that satisfy the equality constraint $\int\delta_1(X)f_1(X)dX=1-\beta$ on the detection probability. Under this equality, minimizing $\bar{\cJ}(\delta_0,\delta_1)$ is equivalent to minimizing the numerator $\int\delta_1(X)\cC_o(X)f_1(X)dX$ in \eqref{J_cost}.

Due to the nonnegativity of $\cC_o(X)$ and our assumption that $\cC_o(X)$ does not contain any atoms we have that \eqref{eq:th1.1} has a unique solution $\lambda_o>0$. Suppose that we are in the case where $\alpha\ge\Pro_0(\cC_o(X)\le\lambda_o)$ and consider a test $\{\delta_0(X),\delta_1(X)\}$ that satisfies the equality $\int\delta_1(X)f_1(X)dX=1-\beta$. We can then write
\begin{multline}
\int\delta_1(X)\cC_o(X)f_1(X)dX-\lambda_o(1-\beta)\\
=\int\delta_1(X)\cC_o(X)f_1(X)dX-\lambda_o\int\delta_1(X)f_1(X)dX\\
=\int\delta_1(X)[\cC_o(X)-\lambda_o]f_1(X)dX\\
\ge\int\indz{\cA}\,[\cC_o(X)-\lambda_o]f_1(X)dX\\
=\int\indz{\cA}\,\cC_o(X)f_1(X)dX-\lambda_o\Pro_1(\cA)\\
=\int\indz{\cA}\,\cC_o(X)f_1(X)dX-\lambda_o(1-\beta),
\end{multline}
where $\cA=\{\cC_o(X)\le\lambda_o\}$.
Comparing the first and the last term yields $\int\delta_1(X)\cC_o(X)f_1(X)dX\ge\int\indz{\cA}\,\cC_o(X)f_1(X)dX$, which proves that \eqref{test1a} is the optimum since it minimizes the estimation criterion and satisfies both constraints. We observe in this case that, for the optimum test, the false alarm constraint can be strict.

Consider now the case $\alpha<\Pro_0(\cC_o(X)\le\lambda_o)$ and let us show that there is a pair $\lambda,\gamma$ for which the test in \eqref{test1b} satisfies both constraints with equality. We are first going to prove that for any $\lambda\ge\lambda_o$ we can find $\gamma(\lambda)\ge0$ to satisfy the equality constraint for the detection probability, namely
\begin{equation}
\Pro_1\left(\frac{f_1(X)}{f_0(X)}[\lambda-\cC_o(X)]\ge \gamma(\lambda)\right)=1-\beta.
\label{eq:app1}
\end{equation}
Call $\psi(\lambda,\gamma)=\Pro_1([f_1(X)/f_0(X)][\lambda-\cC_o(X)]\ge \gamma)-(1-\beta)$, fix $\lambda>\lambda_o$, then we observe that
$\psi(\lambda,0)>\psi(\lambda_o,0)=0$. Furthermore $\lim_{\gamma\to\infty}\psi(\lambda,\gamma)=-(1-\beta)<0$. Consequently there exists $\gamma(\lambda)$ such that \eqref{eq:app1} is true. There are two pairs $\lambda,\gamma(\lambda)$ which we can describe explicitly. From the definition of $\lambda_o$ we know that when $\lambda=\lambda_o$ we have $\gamma(\lambda_o)=0$. Consider now $\lambda\to\infty$ and assume that $\gamma(\lambda)/\lambda\to \bar{\gamma}$, then $\bar{\gamma}$ is the solution to the equation
\begin{equation}
\Pro_1\left(\frac{f_1(X)}{f_0(X)}\ge\bar{\gamma}\right)=1-\beta.
\end{equation}
This is true because the test in \eqref{test1b}, after dividing each side by $\lambda$ and letting $\lambda\to\infty$ reduces to the likelihood ratio test with threshold $\bar{\gamma}$.
Since by assumption we have $\beta>\beta(\alpha)$ where $\beta(\alpha)$ is the probability of miss of the Neyman-Pearson test, we conclude that $\bar{\gamma}>\gamma_{\text{NP}}$. This suggests that
\begin{equation}
\Pro_0\left(\frac{f_1(X)}{f_0(X)}\ge\bar{\gamma}\right)<\Pro_0\left(\frac{f_1(X)}{f_0(X)}\ge \gamma_{\text{NP}}\right)=\alpha
\end{equation}

Now we need to show that there exists a value for $\lambda$ and the corresponding threshold $\gamma(\lambda)$ that satisfy the false alarm constraint with equality, namely
\begin{equation}
\Pro_0\left(\frac{f_1(X)}{f_0(X)}[\lambda-\cC_o(X)]\ge \gamma(\lambda)\right)=\alpha.
\label{eq:app2}
\end{equation}
Call $\phi(\lambda)=\Pro_0([f_1(X)/f_0(X)][\lambda-\cC_o(X)]\ge \gamma(\lambda))-\alpha$. Then, because of our previous analysis, it is easy to verify that $\phi(\lambda)$ has opposite signs for $\lambda=\lambda_o$ and $\lambda\to\infty$, meaning that there exists a $\lambda>\lambda_o$ such that $\phi(\lambda)=0$, or that the false alarm constraint is satisfied with equality.

To show that the test in \eqref{test1b} is optimum, let $\lambda,\gamma(\lambda)$ be the previous pair and consider any test $\{\delta_0(X),\delta_1(X)\}$ that satisfies the equality constraint for the detection probability and the inequality constraint for the false alarm. Then we can write
\begin{multline}
\int\delta_1(X)\cC_o(X)f_1(X)dX-\lambda(1-\beta)+\gamma(\lambda)\alpha\\
\ge\int\delta_1(X)\{[\cC_o(X)-\lambda]f_1(X)+\gamma(\lambda)f_0(X)\}dX\\
\ge\int\indz{\cA}\,\{[\cC_o(X)-\lambda]f_1(X)+\gamma(\lambda)f_0(X)\}dX\\
=\int\indz{\cA}\,\cC_o(X)f_1(X)dX-\lambda(1-\beta)+\gamma(\lambda)\alpha,
\end{multline}
where $\cA=\{\frac{f_1(X)}{f_0(X)}[\lambda-\cC_o(X)]\ge \gamma(\lambda)\}$.
Again comparing the first and the last term, proves optimality of the test in \eqref{test1b} and therefore concludes the proof of Theorem\,\ref{th:1}. \IEEEQED

{\em Proof of Theorem\,\ref{th:2}:}
Due to independence across receivers for the noises $\{w_n(t)\}$ and the reflection coefficients $\{g_{nm}\}$ we deduce
\begin{equation}
\cL(\bar{r}_1,\ldots,\bar{r}_N|\theta)=\prod_{n=1}^N\cL(\bar{r}_n|\theta).
\end{equation}
It is thus sufficient to show that
\begin{equation}
\cL(\bar{r}_n|\theta)=\frac{e^{R_n^H(\theta)(\bQ_n(\theta)+\bI_M)^{-1}R_n(\theta)}}{|\bQ_n(\theta)+\bI_M|}.
\label{eq:app2.1}
\end{equation}
Since $G_n$ is random, we can first compute $\cL(\bar{r}_n|G_n,\theta)$ by conditioning on the coefficients $G_n$ corresponding to the $n$th receiver and then average out $G_n$. For given $G_n$ the received signal $r_n(t)$ under the two hypotheses differs only in the drift, consequently we can apply Girsanov's theorem \cite[Page 191]{Karatzas} to compute the corresponding likelihood ratio. We can treat the complex valued Wiener process $\{w_n(t)\}$ as a two dimensional real valued Wiener process, with the real and imaginary part of the complex process constituting the two independent components of the two dimensional process. Since the corresponding variances, by assumption, are equal to 0.5, it is straightforward to show that
\begin{multline}
 \hfill
\cL(\bar{r}_n|G_n,\theta)=e^{-\int_0^T|G_n^HS_n(t,\theta)|^2dt+2\Real\left(\int_0^T[S^H_n(t,\theta)G_n]dr_n(t)\right)}    %\\%%
=e^{-G_n^H\bQ_n(\theta)G_n+2\Real(R^H_n(\theta)G_n)},
 \hfill
\label{th:girsanov}
\end{multline}
where $\bQ_n(\theta),R_n(\theta)$ are defined in \eqref{eq:th2.2}

In order to compute $\cL(\bar{r}_n|\theta)$ from $\cL(\bar{r}_n|G_n,\theta)$ we need to average out $G_n$.
We recall that the real and imaginary parts of $G_n$ are Gaussian uncorrelated (and thus independent) vectors, each with mean 0 and covariance matrix equal to $0.5\bI_M$. For notational simplicity we drop in all quantities their dependence on $n$ and $\theta$. Let us also define the following decompositions into real and imaginary parts $G=G_r+jG_i$, $R=R_r+jR_i$, $\bQ=\bQ_r+j\bQ_i$ and, finally, denote $\cG=[G_r',G_i']'$, $\cR=[R_r',R_i']'$, $\biQ=[\bQ_r,-\bQ_i;\bQ_i,\bQ_r]$; then we can write the previous likelihood ratio as follows
\begin{equation}
\cL(\bar{r}|G,\theta)=e^{-\cG'\biQ\cG+2\cR'\cG},
\end{equation}
where we used the fact that $\bQ$, by being Hermitian, satisfies $\bQ_r'=\bQ_r$ and $\bQ_i'=-\bQ_i$.
We can now average out $\cG$ by recalling that $\cG\sim\cN(0,0.5\bI_{2M})$. By ``completing the square'' we have
\begin{multline}
\cL(\bar{r}|\theta)=\int e^{-\cG'\biQ\cG+2\cR'\cG}\frac{1}{\pi^M}e^{-\cG'\cG}d\cG\\
=\frac{e^{\cR'(\biQ+\bI_{2M})^{-1}\cR}}{\sqrt{|\biQ+\bI_{2M}|}}\times  %\\%%
\int\frac{e^{-(\cG-(\biQ+\bI_{2M})^{-1}\cR)'(\biQ+\bI_{2M})(\cG-(\biQ+\bI_{2M})^{-1}\cR)}}{\sqrt{\frac{(2\pi)^{2M}}{|2(\biQ+\bI_{2M})|}}}d\cG\\
=\frac{e^{\cR'(\biQ+\bI_{2M})^{-1}\cR}}{\sqrt{|\biQ+\bI_{2M}|}},
\label{eq:app2.3}
\end{multline}
where the last integral is equal to 1 since it is the integral of a Gaussian pdf with mean $(\biQ+\bI_{2M})^{-1}\cR$ and covariance matrix $0.5(\biQ+\bI_{2M})^{-1}$.

From the nonegative definiteness of $\bQ$ we have $Y^H\bQ Y\ge0$ for any complex vector $Y$. Using the observation that for any real vector $Z$, it is true that $Z'\bQ_iZ=0$, as a result of $\bQ_i'=-\bQ_i$, we can show that $[Y_r',Y_i']\biQ[Y_r',Y_i']'=Y^H\bQ Y\ge0$ where $Y=Y_r+jY_i$.  Hence $\biQ$ is nonegative definite as well, implying that $\biQ+\bI_{2M}$ is positive definite. 

Define two square matrices $\bA,\bB$ of size $M\times M$ as the solution to the following two equations: $(\bQ_r+\bI_M)\bA-\bQ_i\bB=\bI_M$ and $(\bQ_r+\bI_M)\bB+\bQ_i\bA=\bz_M$ (there always exists a solution due to the positive definiteness of $\biQ+\bI_{2M}$), then by direct computation we can verify that
$(\biQ+\bI_{2M})^{-1}=[\bA,-\bB;\bB,\bA]$ and $(\bQ+\bI_{M})^{-1}=(\bQ_r+\bI_M+j\bQ_i)^{-1}=\bA+j\bB$. With the help of the previous equalities we have
$\cR'(\biQ+\bI_{2M})^{-1}\cR=R^H(\bQ+\bI_{M})^{-1}R$.
This proves the correctness of the exponential term in \eqref{eq:app2.1}.

What is left to show is that $\sqrt{|\biQ+\bI_{2M}|}=|\bQ+\bI_M|$. Since $\biQ+\bI_{2M}=[\bQ_r+\bI_M,-\bQ_i;\bQ_i,\bQ_r+\bI_M]$, if $\rho$ is an eigenvalue of this matrix with corresponding eigenvector $[Y_r',Y_i']'$ then $\rho$ is a double eigenvalue because by direct computation we can verify that $[-Y_i',Y_r']'$ is a second eigenvector (orthogonal to the first and thus different) for the same eigenvalue $\rho$.  Consequently the $2M$ eigenvalues of $\biQ+\bI_{2M}$ are of the form $\rho_1,\rho_1,\ldots,\rho_M,\rho_M$ with $\rho_n>0$ (because of the positive definiteness of $\biQ+\bI_{2M}$), implying $\sqrt{|\biQ+\bI_{2M}|}=\prod_{n=1}^M\rho_n$. 

We can now verify that if $\rho,[Y_r',Y_i']'$ is an eigenvalue-eigenvector pair of $\biQ+\bI_{2M}$ then $\rho,(Y_r+jY_i)$ is an eigenvalue-eigenvector pair of $\bQ+\bI_M$. This suggests that $\rho,(-Y_i+jY_r)$ must also be an eigenvalue-eigenvector pair for the same matrix. However, we observe that $(-Y_i+jY_r)=j(Y_r+jY_i)$, which means that the two eigenvectors are co-linear and therefore coincide. Consequently for the complex matrix $\bQ+\bI_M$ the eigenvalues are the $\rho_1,\ldots,\rho_M$, meaning that the corresponding determinant satisfies $|\bQ+\bI_M|=\prod_{n=1}^M\rho_n$. This proves the desired equality for the two determinants, demonstrates the validity of \eqref{eq:app2.1} and concludes the proof of Theorem\,\ref{th:2}. \IEEEQED

\end{document}